\documentclass{PoS}
\usepackage{tabularx}
\usepackage[latin1]{inputenc}
\usepackage{pstricks,pst-node,pst-text,pst-3d}
\usepackage{amsmath,amsfonts,amssymb}
\usepackage{subfigure} 
\usepackage{amssymb,fontenc,times,mathptmx,color,graphicx,ifpdf}

\font\openface=msbm10 at10pt 
\newcommand{\be}{\begin{equation}}
\newcommand{\en}{\end{equation}}
\newcommand{\bea}{\begin{eqnarray}}
\newcommand{\ena}{\end{eqnarray}}

\newcommand{\Tr}{\mathrm{Tr}}
\newcommand{\lap}{\mathcal{L}^2}

\usepackage{epsfig}
\PoS{PoS(LAT2005)262}

\title{Simulating the scalar field on the fuzzy sphere }

\ShortTitle{Simulating the scalar field on the fuzzy sphere }

\author{\speaker{Fernando Garc\'{\i}a Flores} \\
        Dublin Institute of Advanced Studies, Centro de
        Investigaci\'on y Estudios Avanzados\\
        E-mail: \email{fergar@stp.dias.ie}}
\author{ Denjoe O'Connor \\         Dublin Institute of Advanced Studies\\
        E-mail: \email{denjoe@stp.dias.ie}}

\author{X. Martin\\
        LMPT, Universit\'e F. Rabelais de Tours \\
        E-mail: \email{xavier@lmpt.univ-tours.fr}}

\abstract{
  The properties of the $\phi^4$ scalar field theory on a fuzzy
  sphere are studied numerically. The fuzzy sphere is a discretization
  of the sphere through matrices in which the
  symmetries of the space are preserved. This model presents three
  different phases: uniform and disordered phases, as in the usual
  commutative scalar field theory, and a non-uniform ordered phase related to
  UV-IR mixing like non-commutative effects. We have determined the
  coexistence lines between phases, their triple point and their scaling.}

\FullConference{XXIIIrd International Symposium on Lattice Field
		 Theory\\ 
		 25-30 July 2005\\
		 Trinity College, Dublin, Ireland}

\begin{document}%

\section{Introduction}
The fuzzy approximation scheme \cite{fuzzy} consists in
approximating the algebra of functions on a manifold with a finite
dimensional matrix algebra instead of discretising
the underlying space as a lattice approximation does.

Here we report our results for a hermitian scalar field
on the fuzzy sphere. We find the collapsed phase diagram 
and in particular we calculate the uniform ordered/non-uniform ordered line 
that was absent in \cite{xavier}.

The current study could be relatively easily repeated for a hermitian
scalar field on other fuzzy spaces. The simplest extension would be to
fuzzy $\mbox{\openface CP}^{\rm N}$. Some variants of the scheme can
be applied to fuzzy versions of $S^3$ and $S^4$\cite{spheres}.  The
study reveals that the non-uniform disordered phase lines should
correspond to a pure matrix model transition.

As an approximation scheme, this ``fuzzification'' is well suited to
numerical simulations of field theories \cite{Nishi}. As a test run,
the first fuzzy approximation to be investigated should be the
simplest one, that of the two dimensional sphere $\mbox{\openface
CP}^1=S^2$.  Both the two--dimensional commutative and Moyal planes 
can be viewed as the limits of a fuzzy sphere of infinite radius.

 \section{The two dimensional $\phi^4$ Model and its fuzzy version}

We are interested in the model:
\begin{equation}\label{eq:accion}
  S[\Phi] = Tr \left[ a \, \Phi ^{\dagger} \left[ L_{i} , \left[ L_{i} ,
  \Phi \right] \right] + b \Phi^{2} + c
  \Phi^{4} \right],
\end{equation}
where $\Phi$ is a Hermitian matrix of size $N$, $b$ and $c$ are mass
and coupling parameters respectively. $L_{i}$ is the angular momentum
generator in the $N$ dimensional unitary irreducible representation of
$SU(2)$.  Since a rescaling of $\Phi$ will allow us to set $a=1$, the
entire phase diagram can be explored by ranging through all real
values of $b$ and positive values of $c$. The conventions of
\cite{xavier} are $a=\frac{4\pi}{N}$, $b=a r R^2$, $c=a\lambda R^2$.

The infinite matrix limit of the action can be taken and corresponds to
a real scalar field $\phi$ on a round sphere of radius $R$
and Euclidean action
\begin{equation}\label{eq:theaction}
  S[\phi] = \int_{S^{2}} d^{2} {\bf n }\left( 
\phi{\cal L}^2 \phi + r R^2 \phi^{2} +  \lambda R^2 
  \phi^{ 4 }  \right)
\end{equation}
where ${\cal L}^2 =\sum_{i=1,3}{\cal L}_{i}^2$ and ${\cal L}_i$ are the
usual angular momentum generators. 

The eigenvectors of $\left[ L_{i} , \left[ L_{i} , \cdot \right]
  \right]$ in (\ref{eq:accion}) are the polarization tensors
$\hat{Y}_{lm}$ (normalised so that $\frac{4\pi}{N}
\Tr(\hat{Y}_{lm}^\dag \hat{Y}_{lm})=1$) 
and it has eigenvalues $l(l+1)$ with degeneracy $2l+1$.
This is precisely the spectrum of the Laplacian $\lap$ on the commutative
sphere truncated at angular momentum $N-1$.

This particular model was chosen because of its simplicity.  The
diagrammatic expansion of the model (\ref{eq:theaction}) has only one
divergent diagram, the tadpole diagram, is Borel resumable, and
defines the field theory entirely. In the fuzzy version, the tadpole
splits into planar and non-planar tadpoles, which are also the only
diagrams that diverge in the infinite $N$ limit. Their difference is
finite and nonlocal and is responsible for the UV/IR
mixing phenomena of the disordered phase \cite{uvirmixing}.

Even though the scalar field on either commutative or fuzzy spheres
cannot have a phase transition, since they have finite volume or a
finite number of degrees of freedom, phase transitions may be found 
when the matrix dimension or the radius of the 
sphere become infinite. 

The fuzzy sphere can be recognized by introducing coordinates
$\left( X_{ 1 }, X_{ 2 },X_{ 3 } \right)$
proportional to the angular momentum operators 
\begin{equation}
  X_{i} = \frac{ 2 R }{ \sqrt{ N^{2} - 1 }} L_{i}.
  \nonumber 
\end{equation}
They must satisfy the algebra
\begin{equation}\label{eq:noncommutative} \
  X^{2}_{1} + X^{2}_{2}+ X^{2}_{3}=R^{2} {\bf
  1}, \hspace{1cm} \left[ X_{i} , X_{j} \right] = i \frac{\Theta}{R} 
  \epsilon_{ijk} X_{k}  \nonumber
\end{equation}
where $\Theta = \frac{ 2 R^2 }{ \sqrt{ N^{2} - 1 }}$ is
the parameter of non-commutativity, $R$ is the radius of the sphere
and ${\bf 1}$ is the unit operator.
The non-commutativity parameter depends on the matrix size,
$N$, and the radius of the sphere, $R$. By taking
different limits we can access different spaces: 
\begin{center}
  \begin{tabular}{||c|c|c|l||}
\hline
    \hspace{0.5cm}$N$\hspace{0.5cm} &\hspace{0.5cm}$R$\hspace{0.5cm}
    & \hspace{0.5cm}$\Theta$\hspace{0.5cm}  & \hspace{0.5cm}Limit
    \hspace{0.5cm} \\
\hline
$N$ & constant = $R$ & $2R^2/\sqrt{N^{2}-1}$ & Fuzzy Sphere \\
$\infty$ & constant = $R$ & $0$ & Commutative Sphere \\
$\infty$ &$\infty$ & $0$ & Commutative plane \\
$\infty$ & $\infty$ & constant = $\theta$ & Moyal Plane \\
\hline
  \end{tabular}
\end{center}

\subsection{Order parameters}

A suitable set of order parameters can be identified from the
coefficients of a mode decomposition in terms of the polarization tensors
basis {\cite{Varshalovich}}
\begin{equation}\label{eq:expansion}
  \Phi = Tr(\Phi) \frac{\bf 1}{N} + \frac{12}{N(N^2-1)}\rho_aL_a +
  \sum_{l=2}^{N-1}\sum_{m=-l}^{+l} c_{lm} \hat{Y}_{lm}.
\end{equation} 
We have separated explicitly $l=0$ and $l=1$ from the expansion
to identify two observables whose expectation values
we used to identify the respective phases. The observables are
$|Tr(\Phi)|$ and 
$\rho^2:=\rho_a\rho_a=\displaystyle\sum_{a=1}^{3}(Tr(L_a\Phi))^2$.
The total power in all coefficients is given by 
$Tr(\Phi^2) = \frac{1}{N}(Tr(\Phi))^2+\frac{12}{N(N^2-1)}\rho^2+
\frac{N}{4\pi} \displaystyle\sum_{l=2}^{N-1}\displaystyle\sum_{m=-l}^{l}
|c_{lm}|^{2}$ and can be used to estimate the importance of the
neglected higher modes.

\subsection{The phases}
\begin{figure}[here]
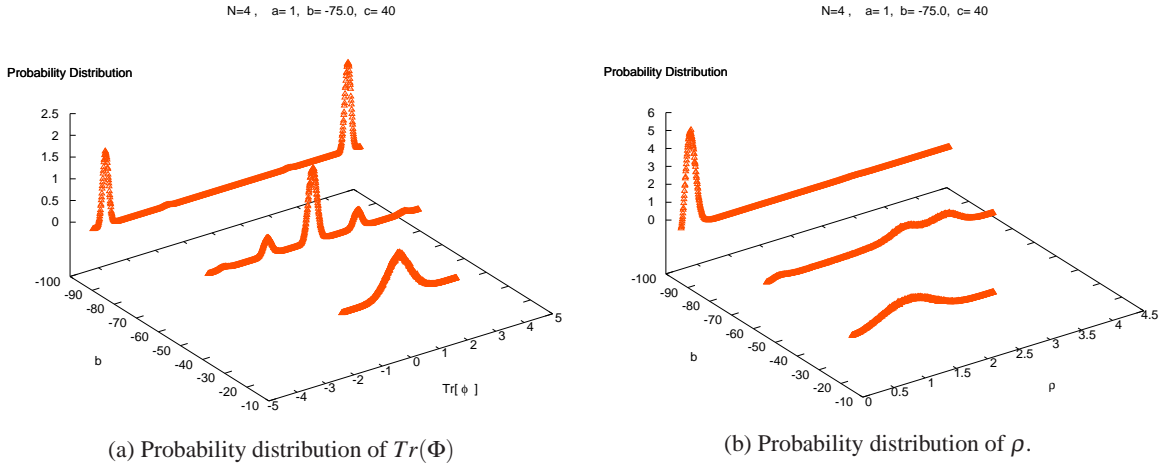

  \begin{center} 
\mbox{ \subfigure[Probability distribution of
  $Tr(\Phi)$]{\scalebox{0.92}{\label{fig:alpha}
  \epsfig{file=histogram3d_4x4_trace.epsi,height=8cm,angle=-90}}}
  \quad \subfigure[Probability distribution of
  $\rho$.]{\scalebox{0.92}{\label{fig:rho}
  \epsfig{file=histogram3d_4x4_rho.epsi,height=8cm,angle=-90}}} }
\caption{Figures (a) and (b) show the typical behavior of the
observables $Tr(\Phi)$ and $\rho$ in a region of the phase diagram
where decreasing $b$ passes the system through the three phases.}
\label{fig:thephases}
\end{center}
\end{figure}

This model (\ref{eq:accion}) presents three phases. As a generic
illustration of their properties, Fig. \ref{fig:thephases}(a) and
Fig.\ref{fig:thephases}(b) show the dependence on the mass parameter
$b$ of the probability distributions of $Tr(\Phi)$ and $\rho$,
respectively, for $\left\{ a=1, \ c=40,\ N=4 \right\}$.

{\bf Disordered:} Found for $|b|$ ``small'', the configurations
  fluctuate close to $\Phi=0$. This is confirmed on the figure,
  $<|Tr(\Phi)|>\sim <\rho>\sim 0$, but also $<Tr(\Phi^2)>\sim 0$ (not
  shown).

{\bf Non-uniform ordered:} As $|b|$ increases, the figure shows
multiple symmetric peaks for the probability distribution of
$Tr(\Phi)$ whose height decreases with increasing $|Tr(\Phi)|$, and
multiple peaks not centered near zero for $\rho$. Furthermore
$Tr(\Phi^2)$ is much larger than both $\frac{<|Tr(\Phi)|>^2}{N}$ and
$<\rho>$ so that higher modes actually dominate. In particular, the
most probable configuration is not rotationally invariant and we have
spontaneous breakdown of rotational invariance. The probability
distribution of $Tr(\Phi)$ has $N+1$ symmetric peaks located
approximately at $(N-2k)\sqrt{-b/2c}$ where $k=0,1,\dots,N$, while the
probability distributions of $\rho$ and $S[\Phi]$ have $(N+1)/2$ peaks
for $N$ odd and $N/2+1$ for $N$ even.

{\bf Uniform ordered:} As $|b|$ becomes large, the figure shows two
  symmetric peaks for the probability distribution of $Tr(\Phi)$
  corresponding to the outer peaks of the non-uniform ordered phase
  and located approximately at $Tr(\Phi)\sim \pm N\sqrt{-b/2 c}$, but
  just one peak near zero for $\rho$. Furthermore, $<Tr(\Phi^2)> \sim
  <Tr(\Phi)>^2/N$ indicating that the power in higher modes is
  negligible. This is generic and indicates that $\Phi\sim
  \sqrt{-b/2c}\,{\bf 1}$ and the rotational symmetry is thus
  restored.
\section{Simulation and Results: The specific heat and the phase diagram}
We are interested in the phase diagram of the model
(\ref{eq:accion}). To identify it, we used the coordinates in
parameter space of the peaks of the specific heat,
$C:=<S^{2}> - <S>^{2}$. The relevant
set of parameters is $\left\{ N , b , c \right\}$ where $b$ and $c$
depend implicitly in $R$. It is possible to further reduce by one the
number of parameters by finding a scaling $\left\{ b , c \right\}
\to \left\{ bN^{\theta_{b}}, cN^{\theta_{c}} \right\}$. If we 
find such $\theta_{b}$ and $\theta_{c}$, the model becomes independent
of $N$ and automatically yields an infinite matrix limit.

\begin{figure}[!h]
\begin{center}
\includegraphics[width=0.53 \textwidth,angle=-90]{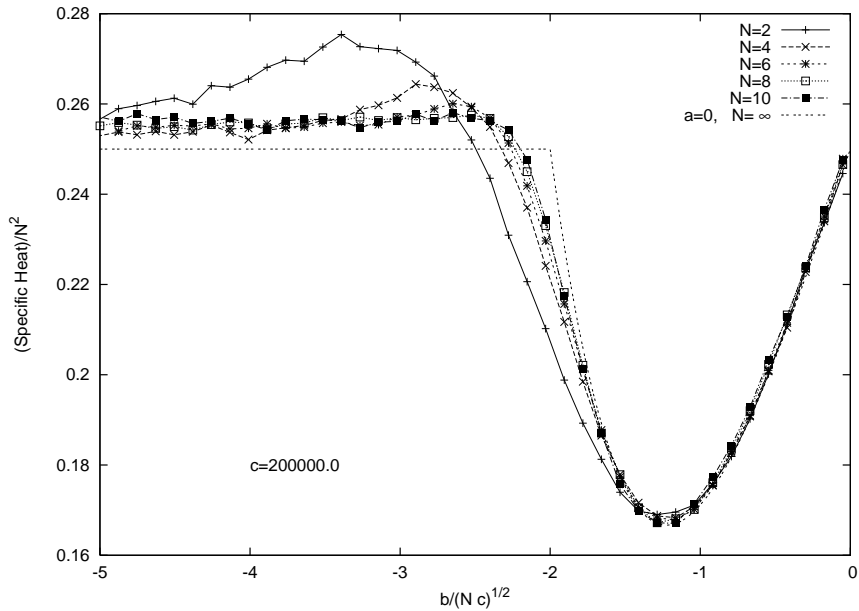}
\caption{Plot of the specific heat at the disordered/non--uniform
  ordered transition for increasing $N$ and its $N\to\infty$
  limit, the exact pure potential model.}
\label{fig:matrix}
\end{center}
\end{figure}

The simulations show that in the non--uniform ordered phase, the fuzzy
kinetic term (proportional to $a$ in (\ref{eq:accion})), is negligible
compared to the potential term (the other terms). There
exists an exact solution for the corresponding limit of $a=0$ in the
large $N$ limit called the pure potential model
\cite{pavelzuberfrancesco}. This model predicts a third order phase
transition between a disordered and non--uniform ordered phase at
$c=b^{2}/4N$. Figure \ref{fig:matrix} confirms numerically the
convergence of the disordered/non--uniform ordered transition towards
this exact critical line of the pure potential model.
\begin{figure}[!h]
\begin{center}
\includegraphics[width=0.53 \textwidth,angle=-90]{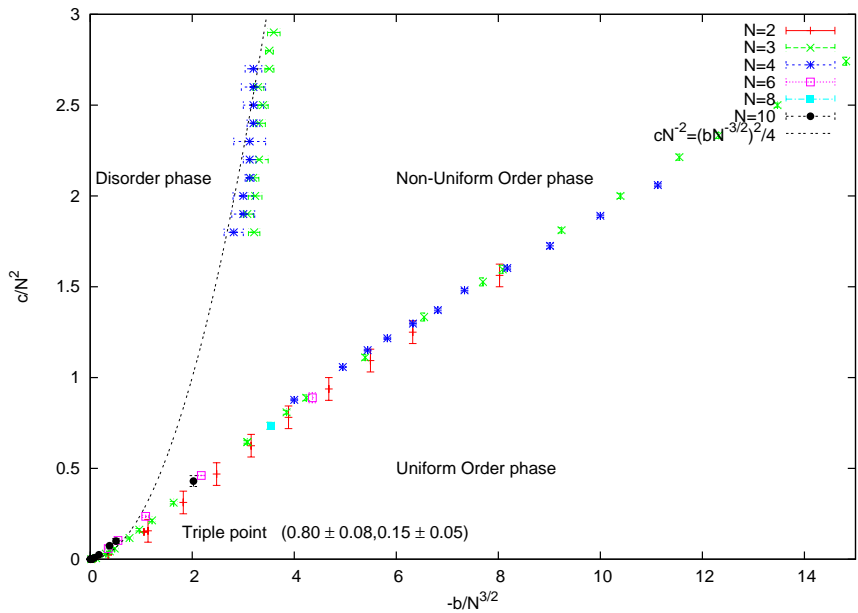}   
\caption{Phase diagram obtained from Monte--Carlo simulations of the
  model (\protect\ref{eq:accion})}
\label{fig:diagram}
\end{center}
\end{figure}

Numerically, it is not difficult to find the coexistence curve between
the uniform ordered and disordered phases which exist for low values
of $c$. On the other hand, the coexistence curve between the two
ordered phases is difficult to evaluate because it involves a jump in
the field configuration and tunnelling over a wide potential barrier.

The phase diagram obtained by Monte Carlo simulations with the
Metropolis algorithm is shown on figure \ref{fig:diagram}. That plot
shows the phase diagram for the model (\ref{eq:accion}). The data have
been collapsed using the scaling form defined above with
$\theta_{b}=-3/2$ and $\theta_{c}=-2$. It is remarkable that this
scaling also works for the exact solution of the pure model potential.

\section{Conclusions}

The numerical study showed three different phases. In one of those
phases, which does not exist in the commutative planar $\lambda
\phi^{4}$ theory, the rotational symmetry is spontaneously broken. The
other two phases have qualitatively the same character as the phases
of this latter model. The three coexistence curves intersect at a
triple point given by
\begin{equation}
(b_{_T},c_{_T})=(-0.15 N^{3/2},0.8 N^{2}).
\end{equation}

Those three curves and the triple point collapse using the same
scaling function of $N$ and thus give a consistent $N
\to \infty$ limit. Thus, all three phases, and in particular the new uniform
disordered phase, and the triple point survive in the limit.
We will discuss these issues more completely in Ref. 
\cite{XavierDenjoeFernando}.

\paragraph{Acknowledgements} We wish to thank W. Bietenholz and J. Medina for
helpful discussions.

\end{document}